\documentclass[conference]{IEEEtran}
\IEEEoverridecommandlockouts

\usepackage{tikz} 
\usepackage{amsmath}
\usepackage{xspace}
\usepackage{enumitem}
\usepackage{cite}
\usepackage{amsmath,amssymb,amsfonts}
\usepackage{algorithmic}
\usepackage{graphicx}
\usepackage{textcomp}
\usepackage{xcolor}
\def\BibTeX{{\rm B\kern-.05em{\sc i\kern-.025em b}\kern-.08em
    T\kern-.1667em\lower.7ex\hbox{E}\kern-.125emX}}

\usepackage{filecontents}
\usepackage{textcomp} 
\usepackage{pifont}
\usepackage{txfonts}
\usepackage{array, boldline, makecell}

\usepackage{booktabs}
\usepackage{caption}
\newcommand{\sys}{Switch$\Delta$\xspace}

\newcommand{\dn}{data node\xspace}
\newcommand{\dns}{data nodes\xspace}
\newcommand{\mn}{metadata node\xspace}
\newcommand{\mns}{metadata nodes\xspace}

\newcommand{\bbproblem}{Ordered Write\xspace}
\newcommand{\bproblem}{Ordered write\xspace}
\newcommand{\problem}{ordered write\xspace}

\newcommand{\mycaption}[3]{{\vspace{-0.4cm}\caption{\label{#1}{#2. }{ #3}}\vspace{-0.25cm}}}
\newcommand{\tablecaption}[3]{{\caption{\label{#1}{#2. }{ #3}}}\vspace{-0.4cm}}
\newcommand{\xxdelta}{Switch$\Delta$}

\usepackage{titlesec}
\titlespacing*{\section}{1pt}{0.5ex}{0.5ex}
\titlespacing*{\subsection}{0pt}{0.1ex}{0.1ex}
\titlespacing*{\subsubsection}{0pt}{0ex}{0ex}

\usepackage{url}
\usepackage{titlesec}
\usepackage{cite}
\usepackage{amsmath,amssymb,amsfonts}
\usepackage{algorithmic}
\usepackage{graphicx}
\usepackage{textcomp}
\usepackage{xcolor}
\def\BibTeX{{\rm B\kern-.05em{\sc i\kern-.025em b}\kern-.08em
    T\kern-.1667em\lower.7ex\hbox{E}\kern-.125emX}}
\begin{document}

\date{}

\title{Switch$\Delta$: Asynchronous Metadata Updating for Distributed Storage with In-Network Data Visibility}


\author{\IEEEauthorblockN{Junru Li} 
\IEEEauthorblockA{\textit{Tsinghua University} \\
Beijing, China \\
lijr19@tsinghua.org.cn} 
\and
\IEEEauthorblockN{Qing Wang} 
\IEEEauthorblockA{\textit{Nanjing University} \\
Nanjing, China \\
wangqing.cs@nju.edu.cn}
\and
\IEEEauthorblockN{Zhe Yang} 
\IEEEauthorblockA{\textit{Tsinghua University} \\
Beijing, China \\
yangzhe.ac@outlook.com}
\and
\IEEEauthorblockN{Shuo Liu} 
\IEEEauthorblockA{\textit{Huawei Technologies Co., Ltd.} \\
Shenzhen, China \\
liushuo15@huawei.com}
\and 
\IEEEauthorblockN{Jiwu Shu}
\hspace{9.5cm} 
\IEEEauthorblockA{\textit{Tsinghua University} \\
Beijing, China \\
shujw@tsinghua.edu.cn} 
\and
\IEEEauthorblockN{Youyou Lu*\thanks{*Youyou Lu is the corresponding
author (luyouyou@tsinghua.edu.cn).}}
\IEEEauthorblockA{\textit{Tsinghua University} \\
Beijing, China \\
luyouyou@tsinghua.edu.cn} 
\and
}

\maketitle


\vspace{-2cm}
\begin{abstract}
Distributed storage systems typically maintain strong consistency between 
data nodes and metadata nodes by adopting \emph{ordered writes}:
1) first installing data; 2) then updating metadata to make data visible.
We propose \sys to accelerate ordered writes by moving metadata updates out of the critical path.
It buffers in-flight metadata updates in programmable switches to enable data visibility in the network and retain strong consistency.
\sys uses a best-effort data plane design to overcome the resource limitation of switches and designs a novel metadata update protocol to exploit the benefits of in-network data visibility.
We evaluate \sys in three distributed in-memory storage systems: log-structured key-value stores, file systems, and secondary indexes.
The evaluation shows that \sys reduces the latency of write operations by up to 52.4\% and boosts the throughput by up to 126.9\% under write-heavy workloads. 
\end{abstract}

\begin{IEEEkeywords}
distributed storage systems, programmable switches, asynchronous metadata updating
\end{IEEEkeywords}

\section{\bf Introduction}


In most distributed storage systems, data and metadata are managed separately to enable flexible and independent resource provision.
This is because the management of data and metadata have inherent divergences in performance goals, functionality requirements, and partition schemes.
For example, distributed file systems~\cite{gfs, PolarFS, ceph,hopsFS, LocoFS, infinifs} spread file data over as many nodes as possible, 
for purposes such as expanding storage space and leveraging aggregated bandwidth.
In contrast, they usually store metadata on a single node or a few nodes, 
to reduce the cross-node synchronization overhead for maintaining the tree-based namespace;
in addition, metadata nodes always require abundant CPU and memory resources, 
to absorb massive metadata requests~\cite{SingularFS,gfs}. 
For another example, when building data stores using shared log systems~\cite{corfu,tango,vCorfu,Delos,ICDElsm1, ICDElsm2},
it is imperative to have an indexing service (which is deployed on dedicated metadata nodes) that maintains the mapping from keys to logs' addresses.


The separation of data and metadata makes read/write operations involve multiple nodes,
requiring distributed coordination.
For read operations (e.g., file read in file systems),
the client first fetches metadata (e.g., inode) from metadata nodes,
and then uses the metadata to obtain data from data nodes.
To ensure strong consistency (i.e., linearizability~\cite{Linearizability}) of storage systems,
write operations adopt the opposite access order: 
installing data to data nodes first, and then updating metadata in metadata nodes.
These two steps are ordered and both are on the critical path:
the storage system can notify the client that the write operation is complete \emph{only after the metadata is updated}, to ensure the visibility of installed data.
Such ordered writes stretch user-perceived latency of write operations,
and have a negative impact on system throughput.

In this paper, we propose \sys, a protocol that accelerates the ordered writes by \emph{making metadata updates asynchronous}.
Therefore, in \sys, a write operation can be committed immediately after installing data (without waiting for metadata updates to complete), enjoying one-RTT latency in the critical path.
However, the main challenge is how to retain strong consistency,
since unfinished metadata updates make committed writes invisible for read operations.

To address the above challenge, 
\sys leverages programmable switches to design an in-switch visibility layer.
Specifically, \sys buffers in-flight metadata updates in the switch and steers read operations to the latest committed data.
For a write operation, when processing the response of data nodes, the switch stores the associated metadata update to the visibility layer. 
Metadata nodes execute metadata updates asynchronously and recycle resources in the switch. 
For a read operation, the switch uses the in-switch metadata as a reply, if the target metadata is in the visibility layer, to guarantee that all committed write operations are visible. 
Thanks to the in-network location of switches, accessing the visibility layer does not induce any extra overhead in the critical path. 

Since programmable switches have limited resources such as memory space,
\sys adopts a \emph{best-effort} design, 
which allows operations to fall back to the original non-accelerated path (i.e., operations executed in two ordered phases) when the switch has insufficient resources.
To guarantee the consistency of the whole system in the presence of the non-accelerated requests, 
\sys uses timestamps to order the concurrent write operations 
of both the accelerated requests (i.e., of which metadata updates are cached in the switch) and the non-accelerated requests.
Further, \sys uses a hash-based design to avoid storing variable-sized keys in the switch,
and leverages a lightweight validation mechanism on the server side to handle the hash collision. 

Asynchronous metadata updates also provide an opportunity for improving metadata update throughput \emph{without increasing the user-perceived latency}.
\sys defers the processing of the metadata requests that are outside the critical path, and 
introduce two efficient \emph{batching mechanisms}, operation combining and prefetching pipeline, 
to improve the throughput of distributed storage systems.

We implement \sys on a Tofino switch~\cite{Tofino}, and integrate it into three distributed in-memory storage systems, including a log-structured KV store, 
a distributed file system based on Octopus~\cite{octopus}, and a distributed secondary index based on SLIK~\cite{slik}. 
\sys accelerates most write operations and relieves throughput pressure on metadata nodes. 
The evaluation shows that \sys reduces the median latency up to 50.0\%, 47.7\%, and 52.4\%, in the three systems, respectively. When data nodes use high-latency replication protocol, \sys still reduces latency by 30.0\%.
Due to the latency reduction and the deferred metadata processing technique, \sys boosts the throughput by up to 126.9\%.  

In summary, we make the following contributions:

\begin{itemize}[leftmargin=*]
    \item \sys, an in-switch data visibility protocol for distributed storage systems, which makes metadata updates asynchronous while retaining strong consistency.
    \item A set of mechanisms to overcome the expressive restriction of programmable switches and a batching mechanism to explore throughput benefits of \sys.
    \item Experiments using three distributed storage systems, 
    demonstrating the performance improvement of \sys.
\end{itemize}

\section{\bf Background and Motivation}
\label{sec:moti}

In this section, we firstly describe the data/metadata separation architecture in distributed storage systems (\S\ref{subsec:moti:separate}), 
and the \textit{ordered write} in this architecture (\S\ref{subsec:moti:orderw}). Then, we analyze the overhead of the ordered write (\S\ref{sub:motieval}) and the opportunities from programmable switches to accelerate it (\S\ref{subsec:moti:switch}).

\subsection{\bf Data/Metadata Separation is Common}
\label{subsec:moti:separate}

Many distributed storage systems adopt an architecture 
where data and metadata are respectively stored in two sets of nodes:  
\dns and \mns.
This architecture supports the flexible resources provision for data and metadata;
they often have different performance and functionality requirements, as well as different partition schemes.
Here, we give three examples using this architecture.


%
\textit{Distributed Log-Structured Key-Value Store.}
Shared log is a popular building block 
to construct distributed applications~\cite{corfu, tango, vCorfu, Delos, ICDElsm1, ICDElsm2}.
it provides simple interfaces to append and retrieve log entries.
In a KV store based on shared log,
the client issues a write operation by appending a log entry containing the KV pair in the log;
for a read operation, the client executes log entries in the log to materialize the KV store.
To avoid redundant execution of log entries for different clients, the KV store uses a dedicated service to maintain an index, 
which maps keys to the address of log entries containing the associated KV pairs. 
In this case, the log and the index are the data and metadata, respectively.

\textit{Distributed File System.}
Lots of distributed file system contains a dedicated metadata service~\cite{gfs, PolarFS, ceph,hopsFS, LocoFS}.
This is because they need to spread file data across many nodes (i.e., \dns) to expand the storage space and leverage the aggregated bandwidth, but spread the metadata in a few metadata nodes (i.e., \mns) to avoid the scalability problem of directory tree management~\cite{fast26ljr,shu2023progress,thdpms}.


\textit{Distributed Secondary Index.}
Secondary index maintains the mapping for primary keys to secondary keys (i.e., a field in objects indexed by the primary key) for storage systems,
to accelerate complex data retrieval.
Some systems adopt \emph{independent partitioning}~\cite{slik, ramcloud,SIIPDPS, kulkarni2017rocksteady}, 
which partitions the secondary index to a set of nodes according to the key range of secondary keys.
This approach can support efficient range operations on both primary keys and secondary keys.
With this partition scheme,
an entry in the secondary index and associated objects are usually \emph{not} on the same node.
In this case, we consider objects (including the primary index) and the secondary index as the data and metadata, respectively.


\begin{figure}[]
  \begin{center}
    \includegraphics[width=\linewidth]{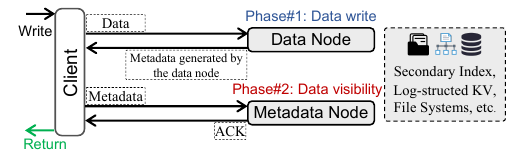}
  \end{center}
  \vspace{-0.1cm}
  \mycaption{pic:moti}{\bproblem}{}
\end{figure}
\begin{table*}[]
    \centering
    \resizebox{1\linewidth}{!}{
    \begin{tabular}{l|l|l|l}
        \toprule
        
         & \multicolumn{1}{c|}{\textbf{Distributed Log-Structured KV}~\cite{corfu, tango, vCorfu, Delos}}   & \multicolumn{1}{c|}{\textbf{Distributed File System}~\cite{gfs, PolarFS, ceph,hopsFS, LocoFS, infinifs}}             &      \multicolumn{1}{c}{\textbf{Distributed Secondary Index}~\cite{slik, ramcloud,SIIPDPS, kulkarni2017rocksteady}}                                                                                  \\ \hlineB{2.5}
         \begin{tabular}[c]{@{}l@{}}\multicolumn{1}{c}{\textsc{\textbf{Write}}} \\ \multicolumn{1}{c}{\textsc{\textbf{Interface}}}\end{tabular}&  \texttt{put(key, value)}      & \begin{tabular}[c]{@{}l@{}}\texttt{fd = open(pathName)} \\ \texttt{write(fd, buf, cnt)}    \end{tabular}  & \begin{tabular}[c]{@{}l@{}}\texttt{put(pKey, value, sKey, ...)} \\ \textit{pKey: primary key, sKey: secondary key}  \end{tabular}  \\ \hlineB{2.5}
        \multicolumn{1}{c|}{\textsc{\textbf{Data Node} }}    &  \texttt{logID \texttt{--\textgreater{}} key-value pair}                & \texttt{blockID \texttt{--\textgreater{}} data block} &   \texttt{pKey \texttt{--\textgreater{}} value}       \\ \hline
        \multicolumn{1}{c|}{\textsc{\textbf{Metadata Node} }}   & \texttt{key \texttt{--\textgreater{}} logID}   & \texttt{pathName \texttt{--\textgreater{}} inode}      &  \texttt{sKey \texttt{--\textgreater{}} [pKeys]}                                 \\ \hline

        \multicolumn{1}{c|}{\textsc{\textbf{Ordered Write}}  }   & \begin{tabular}[c]{@{}l@{}} \texttt{\ding{182} logID = write(key-value pair)} \\ \texttt{\ding{183} update(key, logID)} \end{tabular} &     \begin{tabular}[c]{@{}l@{}}\texttt{\ding{182} block\_list = write(buf, cnt)} \\ \texttt{\ding{183} update(fd, block\_list)} \end{tabular}         &  \begin{tabular}[c]{@{}l@{}}\texttt{\ding{182} update(pkey, value)}     \\ \texttt{\ding{183} update(skey, pkey)} \end{tabular}      \\ \bottomrule
    \end{tabular}
    }
    \vspace{-0.1cm}
    \tablecaption{tbl:orderw}{\bproblem in different storage systems}{\ding{182} Data write phase, \ding{183} Data visibility phase.}
\end{table*}



\subsection{\bf Consistency with Ordered Write}
\label{subsec:moti:orderw}
Since the metadata and data are located on different nodes,
the write and read operations will involve more than one node, 
inducing distributed coordination.
To provide strong consistency (i.e., linearizability~\cite{Linearizability}),
storage systems need an approach to ensure that   
\emph{a read operation will see the atomic effects of all write operations that finished before it started.}

It is straightforward to achieve this using distributed transactions.
However, distributed transactions have poor performance which results from distributed concurrency control protocol.
Many distributed storage systems use \textit{\problem} to achieve strong consistency, considering that 
the storage operations usually only involve one data node and one metadata node.
As shown in Fig.\ref{pic:moti}.a, the write operation has two ordered phases before returning:

\textit{1) Data Write Phase}. The client sends the data to a data node. The data node installs the data and returns a response, which may contain a metadata record generated by it.

\textit{2) Data Visibility Phase}. The client sends the metadata record to a metadata node.
The metadata node installs the metadata record and responds.
After that, any client can fetch the metadata record from the metadata node, and then obtain the data from the data node using the metadata record.
In other words, the data is \emph{visible} to all clients after this phase, guaranteeing linearizability.

Table~\ref{tbl:orderw} provides a summary of the ordered write process in different storage systems. For instance, in a log-structured KV store, the data write phase involves appending to the log and getting a log address, while the visibility phase updates the key-to-address index. Similarly, a file system writes data blocks before updating the inode, and a secondary index system updates the primary object before making the change visible in the secondary index.






\subsection{\bf \bbproblem Overhead}
\label{sub:motieval}

The ordered write ensures strong consistency for storage systems,
but has performance limitations in terms of latency and throughput.
First, it executes two phases (i.e., two RPCs) sequentially in the critical path, 
posing a negative impact on user-perceived latency. 
Second, the processing capacity of metadata nodes and data nodes varies under different workloads at different times.
By stringing operations in metadata nodes and data nodes, 
\problem results in the application throughput being limited by the minimum processing capacity of metadata nodes and data nodes.
In this work, we ask a question:
while \problem seems to be the simplest way to ensure strong consistency,
can we exploit the recent hardware advancement to accelerate it?

\subsection{\bf Opportunities from Programmable Switches}

\label{subsec:moti:switch}
  
Ordered write inherently involves two RPCs, since the data visibility phase relies on the metadata generated by the data write phase.
To shorten the user-perceived latency, 
storage systems need to move the data visibility phase outside the critical path,
i.e., notifying upper applications that a write operation has been completed 
\emph{immediately after the data write phase has finished}, and making the data visibility phase \emph{asynchronous}.
Therefore, we need a location (called \emph{data visibility layer}) to record in-flight metadata updates,
to enable read operations to correctly find the latest data.
We argue that emerging in-network acceleration techniques~\cite{switchtx,p4db,1pipe,netcache,netlock,Kim1,AlNiCo,kim2,kim3} have the opportunity to build this layer, 
since the states in programmable switches or NICs can be accessed in the path of network transmission. 
In this work, we focus on programmable switches due to they are located in the common path for all clients and storage nodes.




Specifically, programmable switches support user-defined packet processing~\cite{netcache, netchain,farreach,distcache}.
They also have on-chip memory that can be read and written in the form of \emph{registers} (fixed-length 8/16/32/64 bit arrays).
When a packet arrives, the switch first parses the packet header.
After that, the switch matches the table entry in each \emph{match-action table} with the packet header and then executes the associated action.
The action can modify on-chip memory and packet header with simple comparisons and calculations.
Current programmable switches have constrained on-chip memory capacity (e.g., $\sim$10MB in Intel Tofino~\cite{Tofino} and 20$\sim$64MB in Arista 7170 Series~\cite{Arista}) and limited payload parsing capability. 

\section{\bf \sys Design}
We propose \xxdelta, an in-network data visibility protocol designed to separate the data write phase and the data visibility phase of \problem. This aims to enhance the performance of distributed storage systems. The \textbf{key idea} of \sys is to store in-flight metadata updates in the switch, ensuring that data becomes immediately visible after the data write phase. Simultaneously, these metadata updates are applied to metadata nodes asynchronously.

Moreover, moving the data visibility phase outside the critical path provides a chance to improve throughput via efficient batching processing 
\emph{without increasing the user-perceived latency}:
we can accumulate a batch of asynchronous metadata updates and execute them at once. 
For example, inserting multiple items into a tree index can reuse the results of tree traversal and have good cache locality~\cite{corobase}.



We begin by presenting the in-network data visibility protocol in \S\ref{sub:protocol} and \S\ref{sub:CC}. Subsequently, we describe the delayed metadata processing design, which enhances the advantages of the in-network data visibility protocol in terms of throughput, discussed in \S\ref{sub:batch}. Lastly, 
we outline the strategies for handling node or switch failures in \S\ref{sub:failure}.

Throughout this section, we employ log-structured in-memory key-value (KV) stores to illustrate \sys, as they are simple yet representative of the ordered write problem. 
In \S\ref{sec:app}, we elaborate on how to integrate \sys into distributed file systems and distributed secondary index systems.

\subsection{\bf Protocol Overview}
\label{sub:protocol}


In the \sys cluster, nodes communicate with each other through Remote Procedure Calls (RPCs), with each RPC packet augmented by a \sys header. \sys employs an in-switch hash table as its visibility layer. The hash table uses the hash value of the queried key as an index. Each entry in the hash table stores in-flight metadata updates, which map the key to the address of its newly allocated log entry. These updates are stored before being applied to \mns, and are subsequently used to service read operations for strong consistency. 
Depending on the type of RPC, \sys is capable of installing/deleting/searching a metadata entry in the hash table based on the packet.


Here, we initially outline the workflow of write/read operations and their interaction with the in-switch hash table, disregarding concurrent write operations and the hash collision problem of the hash table. Subsequently, we delve into the concurrency control mechanism in \S\ref{sub:CC}.


 
\noindent
\textit{1) Write Operations.}
Fig.\ref{pic:read}-a illustrates the write operation workflow in \sys, which completes in a single round-trip on the critical path. For a write operation, the client first initiates a data write request (\ding{182}\ding{183}) to the data node (\dn). The \dn stores the KV pair in an allocated log entry and returns a reply containing the new metadata (the key and log entry ID) (\ding{184}).

For comparison, in the original system (Fig.\ref{pic:moti}), the client would receive this reply and then send a separate metadata update request to the metadata node. The operation would be considered complete only after receiving confirmation from the \mn.

\sys streamlines this process. Upon receiving the reply from the \dn, the switch immediately stores the metadata, making the write operation visible. It then forwards the reply to both the client (\ding{185}) and, asynchronously, to the \mn (\ding{175}). Once the client receives its copy of the reply (\ding{185}), the operation is considered committed and is acknowledged to the application. This design consolidates the data writing and visibility-granting steps into a single round-trip. 
In the background, the \mn processes its copy of the reply (\ding{175}) by applying the update to its index. Afterwards, it instructs the switch to clear the corresponding cached metadata (\ding{176}). 




\begin{figure}[]
  \begin{center}
    \includegraphics[width=\linewidth]{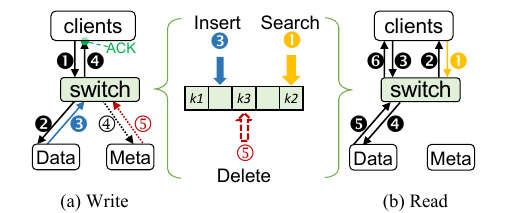}
  \end{center}
  \mycaption{pic:read}{Workflows of write and read operations}{}
\end{figure}


\noindent
\textit{2) Read Operations.}
\label{subsub:read}
\sys divides the metadata into two parts: a small and newer portion in the switch and a larger portion in \mns. Consequently, as shown in Fig.\ref{pic:read}.b, read operations must initially search for metadata in the switch before searching in \mns. For each read operation (\ding{182}), if the switch determines that the queried metadata is stored within its domain, it copies this metadata entry to the request and routes it as the reply to the client (\ding{183}); otherwise, the read operation falls back to the original path.

\subsection{\bf Concurrency Control}
\label{sub:CC}
To manage concurrent writes and adapt to switch hardware limitations, \sys incorporates two primary mechanisms.
\textbf{The first} is a timestamp-based protocol to handle concurrent writes to the same key. This guarantees that the metadata cached in the switch corresponds to the most recent successfully committed write.
\textbf{The second} mechanism addresses the challenge of large and variable-length keys. Instead of using raw keys, \sys indexes in-switch metadata using their hash values. This design is friendly to the switch's limited resources but necessitates a strategy for handling hash collisions, which we will now discuss.


\noindent
\textit{\bf 1) Handling Conflicting Write Operations.} 
\label{subsubsec:conflict}
The \dn assigns a timestamp to each log entry before storing new log, and it encodes this timestamp into the reply message (\ding{184}). \sys relies on timestamps to determine the order of write operations to the same in-switch metadata entry.
Therefore, the keys with the identical hash value are required to utilize the same timestamp generator. 
To avoid remote timestamp synchronization, \sys imposes a requirement on the data partition scheme: data associated with keys sharing the same hash value must be stored within the same \dn, as illustrated in Fig.\ref{pic:timestamp}. Essentially, each metadata entry in the switch is linked to a specific \dn.

%
Each entry in the in-switch hash table contains two timestamp registers: \texttt{CurTs} and \texttt{MaxTs}. The \texttt{CurTs} register represents the timestamp of the metadata stored in the entry, while the \texttt{MaxTs} register indicates the maximum timestamp among the write operations that have utilized the same entry.
%

\begin{figure}[]
  \begin{center}
    \includegraphics[width=\linewidth]{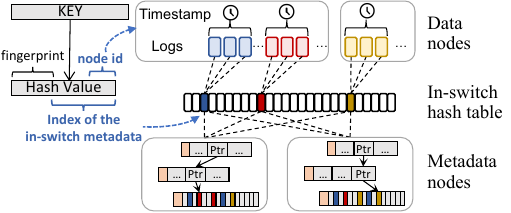}
  \end{center}
  \mycaption{pic:timestamp}{Timestamp generators and the partition scheme}{}
\end{figure}

When the switch receives a data write reply from the \dn and attempts to store metadata in the hash table (\ding{184}), it compares the timestamp of the message with the \texttt{MaxTs} register of the corresponding entry. It stores the metadata only if the entry is clear, and the timestamp of the message is larger than that of the entry. It is important to note that \sys does not permit an operation with a higher timestamp to overwrite an older one. This restriction is in place because allowing the overwriting mechanism would lead to inaccuracies in the event of packet loss. If a write operation discovers that its metadata entry is not clear, the operation falls back to the original path (i.e., two phases in ordered writes).

When the switch receives a metadata update reply from the \mn and attempts to clear the metadata entry (\ding{176}), it compares the timestamp of the request with the \texttt{CurTs} register of the corresponding entry. It clears the metadata entry only if these two timestamps are equal. Due to packet reordering in the network, before receiving the reply with the equal timestamp, the switch may receive the metadata update reply of another operation that has a higher timestamp. The switch blocks returning this operation to clients and allows the \mn to re-send this reply until the metadata entry is cleared. This ensures that the metadata in the switch represents the newest version of committed operations.

\noindent
\textit{\bf 2) Handling In-Switch Hash Collision.}
\label{subsubsec:hashcollision}
The hash value size is fixed and usually smaller than the key size. Consequently, two different keys can have the same hash value and share the same metadata entry in the switch.

\textit{For write operations with the same hash value},  
\sys does not attempt to find another slot to store the metadata that caused the collision. If a write operation discovers that its metadata entry is not clear, this operation falls back to the original path (i.e., two phases in ordered writes).

\textit{For read operations},
allowing different operations to share the same metadata entry poses a challenge for subsequent read operations, as they cannot confirm whether the metadata returned by the switch belongs to their queried key. Therefore, \sys introduces an additional validation stage when the \dn processes the data read request. The \dn stores the entire key with the log entry and checks whether the key in the log entry is equal to the queried key before replying to clients. If a validation failure occurs, the client retries the operation. It's important to note that the client may need to retry multiple times until either: 1) the metadata entry in the switch corresponds to the queried key; or 2) the incorrect metadata entry in the switch is cleared, and this operation retrieves metadata from the \mn.

To minimize the probability of hash collision, each metadata entry also retains the key's fingerprint. If the key size is smaller than the combined size of the hash index and fingerprint 
(16 bits and 32 bits, respectively), no hash collision occurs. For larger key sizes, the validation mechanism ensures the correctness of read operations.



\begin{figure}[]
  \begin{center}
    \includegraphics[width=\linewidth]{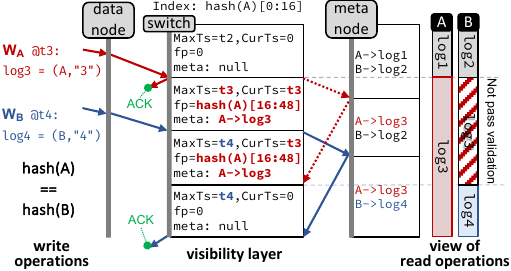}
    \mycaption{pic:example}{Writes with different keys but the same hash value}{}
  \end{center}
\end{figure}




\noindent
{\bf Example of Hash Collision.}
Fig.\ref{pic:example} illustrates a \textbf{corner case} involving two concurrent write operations,  $W_A$ on \texttt{A} and $W_B$ on \texttt{B}, each with different keys but the same hash value (i.e., \texttt{hash(A) == hash(B)}). 
Initially, the value of \texttt{A} is stored in $\texttt{log}_\texttt{1}$, and the value of \texttt{B} is stored in $\texttt{log}_\texttt{2}$. 
This example omits the packets from the client to the \dn.
$W_A$ gets the timestamp $\texttt{t}_\texttt{3}$ and writes the new value to $\texttt{log}_\texttt{3}$. 
When the reply of the data write arrives at the switch, it stores metadata in switch, making the metadata $\texttt{A}\rightarrow\texttt{log}_\texttt{3}$ visible. $W_A$ is the finished in one round trip.

$W_B$ represents operations falling back to the original path due to hash collision. It obtains timestamp $\texttt{t}_\texttt{4}$ ($\texttt{t}_\texttt{4}>\texttt{t}_\texttt{3}$) and writes the new values of \texttt{B} to $\texttt{log}_\texttt{4}$. When the reply of the data write phase arrives at the switch, the switch finds the corresponding entry is not empty, causing $W_B$ to fall back to the original path. After the data visibility phase of $W_A$, the switch clears this metadata entry, enabling subsequent write operations on \texttt{B} to be accelerated.

The right part of Fig.\ref{pic:example} shows the view of read operations on \texttt{A} and \texttt{B}. During the background metadata updates of $W_A$, read operations on \texttt{B} cannot pass the validation step in the \dn and are blocked until the metadata entry is cleared.

This example illustrates why metadata with the same hash value \textit{cannot} overwrite each other. If the request of $W_A$ sent from the switch to the metadata node is lost and the subsequent $W_B$ were to overwrite the in-switch metadata of $W_A$, and neither the switch nor the metadata node has the metadata: $\texttt{A}\rightarrow\texttt{log}_\texttt{3}$. In this scenario, the system would be unable to detect the lost packet. $W_A$ has already been committed, but the subsequent read operations cannot read the corresponding metadata, resulting in inconsistency.

\subsection{\bf Partial-Write Operations}
\label{sub:pw}

In the description of \sys protocol, we assume that each write operation updates an entire metadata entry, which is referred as \textit{full-write} (FW) operations. 
Some storage systems use the \textit{partial-write} (PW) operations: a partial-write operation only updates a part of its metadata entry, such as a file write only updates the inodes of the modified file range. 

In this system using PW, the switch stores the updated part of the metadata entry. 
Therefore, a metadata entry is partitioned among the switch and the \mn. 
\sys employs a different design for the read operations. 
Specifically,  
when the switch receives the read request, it adds the partial metadata entry stored in the switch and routes this new request to the \mn. The \mn applies the partial metadata entry to the metadata and then sends the entire metadata entry to the client.

\subsection{\bf Deferred Metadata Processing}
\label{sub:batch}

With \sys, nearly all metadata update requests are outside the critical path. To further exploit its benefits, \sys employs a deferred metadata processing method. This method serves a dual purpose: allowing the CPU to focus on critical path requests such as non-accelerated metadata update and query requests, and grouping deferred requests into batches to leverage throughput-enhancing batching techniques.

A \mn promptly processes requests on the critical path and temporarily buffers asynchronous metadata requests. When the metadata node observes that the number of deferred requests exceeds the buffer size or there are no other requests in progress, it starts to process the deferred requests in batches. \sys introduces two batching techniques: operation combining and prefetching pipeline.

The first technique is \textit{operation combining}.
In tree-based indexing, performing update/insert operations in sorted order according to the keys enhances throughput by improving cache locality, as neighboring operations involve adjacent nodes. Therefore, \sys sorts the requests for each batch.

The second technique is \textit{prefetching pipeline}.
For each operation, especially on the tree-based index, the CPU accesses the tree nodes randomly, which is expensive when they are not cached. \sys adopts a similar idea to CoroBase~\cite{corobase} to reduce the overhead. \sys uses coroutines to pipeline the memory prefetch and the CPU processing. For each batch, \sys initiates a certain number of coroutines and assigns an operation to each of them. When the coroutine needs to access a tree node (i.e., random memory access), the coroutine issues a prefetch instruction and switches to another coroutine. When the coroutine completes an operation, \sys switches the coroutine for the next operation in the batch.

\sys schedules the coroutines in a round-robin manner. Since L3 cache misses cost about 100ns, when the CPU switches back to the coroutine, the prefetched tree node is already in the cache. Furthermore, since the CPU stall time for L3 cache misses is much larger than the twice time of the coroutine switch (about $8ns\times 2$ in our environment), this optimization can improve throughput (via saving CPU stall cycles) when the tree node is not in the cache.

Note that operation combining and the prefetching pipeline are two general batching optimizations for storage systems. Systems can employ other application-specific or hardware-specific batching designs to further explore the benefits of disentangling the data write phase and data visibility phase. For example, some NICs support the doorbell batching optimization~\cite{guidelines}, which reduces CPU-generated MMIOs from the batching size to one. Exclusive deferring and batch processing of non-critical path operations improve throughput without concerns about latency overhead.

\subsection{\bf Failure Handling}
\label{sub:failure}
The in-switch design introduces new failure scenarios. This section outlines our strategies for ensuring system robustness against packet loss, node crashes, and switch failures.

\noindent
\textit{1) Packet Loss.}
Our system handles packet loss events using timeout-based mechanisms, with different procedures for the packets in critical paths non-critical paths.

Loss of a packet on the critical path results in a client-side request timeout, which prompts a retry for the affected operation. Conversely, the loss of a non-critical path packet, such as a metadata update acknowledgement from the server, leaves a stale entry in the switch. This temporarily prevents subsequent operations mapping to the same hash entry from being accelerated. The metadata node detects this situation after a timeout by not receiving the expected acknowledgement and then issues a command to invalidate the stale switch entry. 

We configure this timeout to 500$\mu s$, a value approximately 100 times the typical round-trip time, making it highly probable that a timeout indicates a genuine packet loss rather than mere network jitter. 
Crucially, all operations are designed to be idempotent, which makes these retries safe. Client requests include a unique request ID, allowing the server to detect and correctly handle duplicate submissions. Similarly, the reclaim messages from the \mn include timestamps, ensuring that the cleanup process is also safe and does not incorrectly invalidate a newer, valid entry.

\noindent
\textit{2) Node Crash.}
Our design's fault tolerance logic is primarily concerned with the metadata node. Failures of data nodes (\dns{}) are handled transparently by the underlying storage system's native recovery protocol, such as data replication. \sys{} does not alter this process. In the event of a metadata node (\mn{}) crash, a new \mn{} instance is initialized. The \dns{} then play a crucial role in recovery: since they store metadata encoded within the data itself, they extract and replay this metadata to the new \mn. This ensures that even in-flight write operations committed in the switch are not lost.

\noindent
\textit{3) Switch Crash.}
A switch crash results in the immediate loss of all in-flight metadata stored in its hash table. To ensure consistency, the system enters a coordinated recovery mode. First, all \mns{} pause and drain their existing request queues. Then, to guard against data loss from concurrent failures (e.g., a lost packet during the crash), the \mns{} perform a synchronization step with all \dns{} to retrieve the state of the most recently committed operations. 

Table~\ref{tbl:failure} outlines our recovery strategies for various failure scenarios. As shown, the system's response is tailored to the failure type. Packet losses, whether on the critical or non-critical path, are handled with low-latency, localized mechanisms. Node crashes are more impactful, with recovery for a metadata node (\mn{}) involving a full state rebuild, while a data node (\dn{}) crash defers to the underlying replication protocol. A switch crash has a global impact, temporarily disabling all acceleration features pending a system-wide recovery.
\begin{table*}[hbt!]
  \centering
  
  \resizebox{0.95\textwidth}{!}{%
  \begin{tabular}{l l l l}
    \toprule
    \textbf{Failure Scenario} & \textbf{Recovery Strategy} & \textbf{Recovery Time} & \textbf{Scope of Impact} \\ 
    \midrule
    Packet Loss (Critical Path) & Client-side timeout and operation retry. & Low (ms) & Small (affects only the single operation). \\ 
    Packet Loss (Non-Critical Path) & Metadata node cleans up the stale in-switch entry. & Low (ms) & Small (blocks same-key operations). \\ 
    \midrule
    Node Crash (\mn{}) & Rebuilds lost state based all \dns{}. & High (second) & Large. \\ 
    Node Crash (\dn{}) & High-availability protocols in original systems. & - & - \\ 
    \midrule
    Switch Crash & Switch reboots; system-wide state resynchronization. & High (sec-min) & Large (all in-network state is lost). \\
    \bottomrule
  \end{tabular}%
  }
  \vspace{-0.1cm}
  \tablecaption{tbl:failure}{Overheads of Failure Handling and Recovery}{}
\end{table*}

\section{\bf Implementation and Deployment}
\label{rsec:impl}

We first describe the packet header of metadata messages in \sys protocol and then describe the switch data plane.

\subsection{\bf Implementation Details}
\label{rsec:impl1}
\begin{figure}[]
  \begin{center}
    \includegraphics[width=\linewidth]{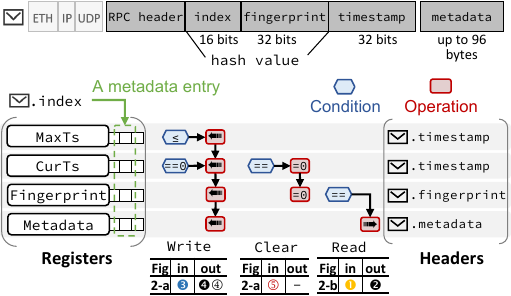}
  \end{center}
  \mycaption{pic:vt}{Details of switch data plane}{The circled numbers in the description match the steps illustrated in Fig.\ref{pic:read}.}

\end{figure}

\textit{1) Network Stack and Packet Header.}
Our RPC framework is built on SEND/RECV verbs of the ibverbs library using RoCEv2 protocol.
Specifically, the data messages that do not need to be modified by the \sys use the reliable connection (RC) transport, and the metadata messages use the RAW PACKET transport. 
The packet format is shown at the top of Fig.\ref{pic:vt}.
We use the UDP source port to identify whether a message uses in-switch data visibility protocol and use the UDP destination port to identify which receive queue (RQ) the message belongs to. \sys header is located after the UDP header.
Besides the source, destination and operation type of RPC, the header also contains the hash index, the fingerprint of the key and the timestamp of the operation.



%

\noindent
\textit{2) Switch Data Plane.}
As shown in Fig.\ref{pic:vt}, the switch provides 3 functions: write, read, and clear.
They have some trigger conditions, such as checking the fingerprint and the timestamp. 
The conditions are number comparisons (i.e., $<=>$) between the packet fields and the corresponding registers in the switch.
In detail, the switch stores metadata entries in a hash table. 
In addition to the metadata itself, each entry contains a fingerprint, a \texttt{MaxTs} register, and a \texttt{CurTs} register.  
in-switch functions include three steps: 
1) finding the entry via the hash value, 
2) comparing fingerprints and timestamps between the packet and the entry, 
3) executing read/write/clear if this operation satisfies all conditions. 


\subsection{\bf Hardware Limitations and Resource Usage}
\label{rsec:impl2}
There are three hardware limitations in our data plane design. First, due to limited on-chip memory, the number of entries in the switch is restricted to $2^{16}$. Second, owing to the limited programmability of in-switch comparisons, the compared field is limited to 32 bits. Therefore, the hash value consists of 48 bits (16-bit index and 32-bit fingerprint), and the timestamp is 32 bits. Additionally, due to the restricted payload parsing capability of our current hardware, the size of a metadata entry is capped at 96 bytes. 

\sys~is lightweight in its use of the switch's limited on-chip memory. The in-network visibility layer only caches metadata for in-flight updates, and entries are evicted upon completion. Each cache entry is compact, consisting of a 48-bit hash value and a timestamp. This design ensures that memory consumption is proportional to the number of concurrent operations, not the dataset size, making it practical for large-scale systems.



\subsection{\bf Discussion on Practicality and Generality.}
\label{rsec:impl3}
In this subsection, we discuss the practical considerations for deploying \sys~in modern data centers.

\begin{itemize}
    [itemsep=0pt, parsep=0pt, labelsep=5pt, leftmargin=*, topsep=0pt, partopsep=0pt]
    \item {\textit{Effectiveness in Diverse Hardware/Software Environments}: While \sys~shows significant relative gains in low-latency settings (e.g., RDMA + memory), its absolute latency reduction is even more impactful in high-latency environments. By eliminating a full RPC round trip, the time saved becomes more pronounced when using slower storage or kernel-based networking stacks (e.g., TCP), where each round trip incurs substantial overhead. Similarly, the performance benefits of \sys~are more pronounced for smaller file or block sizes where network latency dominates the operation time. For operations involving very large blocks, the data transfer time can overshadow the latency savings from eliminating an RPC. In such cases, the relative improvement is smaller because the saved time is a smaller fraction of the overall operation duration. }
    \item {\textit{Systems based on One-sided RDMA}}: Our system can not be used to support systems based on one-sided RDMA, such as FaRM~\cite{farm}. This is because one-sided RDMA protocol has complex hardware-based reliable mechanisms and ack-batching mechanism, which makes it difficult to modify packets in the programmable switch. Compared to one-sided RDMA, RPC-based distributed systems are flexible and we can use the programmable switch to accelerate the data path of RPC while maintaining complex reliable mechanisms in the software.
    \item {\textit{Deployment and Cost}}:
    Deploying \sys~involves loading a P4 program onto the switch. While programmable switches have a higher per-port cost than commodity switches, their adoption by major cloud providers~\cite{alibaba,facebookswitch} indicates a clear industry trend. 
    \item \textit{Environments with Multiple Racks}: In environments with multiple racks, each rack's Top-of-Rack (ToR) switch is configured as the data visibility layer for all metadata nodes within that rack. Workloads are partitioned among these ToR switches, and each switch manages the metadata associated with its partition (i.e., metadata of \mns in its rack).
\end{itemize}

\section{\bf Evaluation}

We evaluate how \sys contribute to the performance of a log-structured KV store.
After describing the experimental setup in \S\ref{exsetup}, we first evaluate the overall performance of \sys to show the benefits of in-network visibility and deferred metadata processing, respectively, in \S\ref{sub:overall}. 
Then we evaluate the performance under various configurations to further analyze the scenarios for which \sys is applicable in \S\ref{sub:sensitive}. Then we evaluate the performance of \sys in a configuration with 3-way data replication in \S\ref{sub:rep}. Finally, we evaluate in depth the contributions and limitations of each technique to the performance improvement in \S\ref{sub:indepth}.

\subsection{\bf Experimental Setup}
\label{exsetup}

\noindent
\textit{1) Hardware.}
We run the experiments with 16 nodes (5 data nodes, 5 metadata nodes and 6 client nodes) by default unless otherwise noted. 
Each node is equipped with a 12-core Intel Xeon E5-2650 v4 2.20GHz CPU, memory of 64GB and one 100Gbps Mellanox ConnectX-5 NIC. 
They are connected by a Barefoot Tofino Wedge 100BF-32X switch. For the baseline, this same switch was configured to act purely as a forwarding device. This methodology ensured a fair comparison by isolating the performance gains attributable solely to our in-switch logic.

\noindent
\textit{2) Thread Model.}
The threads in our experiments are polling-based and bound to unique CPU cores. 
By default, each data node and metadata node uses 4 threads. Each client node uses 8 threads to initiate operations, and the queue depth of each thread is 8 (6 $\times$ 8 $\times$ 8 = \texttt{384} concurrent operations). 
We show the latency against the throughput by varying the number of client threads and the queue depth of each thread.

\noindent
\textit{3) Distributed Log-Structured KV Store Configurations.}
We implement an in-memory log manager in \dns, 
where the logID of each log entry is 32 bits and the size of each record is 128 bytes, the key is 8 bytes. 
The key space is 250 million, and the keys are pre-generated randomly. The key distribution follows the Zipf distribution~\cite{breslau1999web} with the parameter $\theta$=0.99, where 49.1\% operations access the hottest 1\textpertenthousand~ keys with the configuration of 250M key space. The probability of at least one hash collision occurring among 1024 concurrent requests with a 48-bit hash value is extremely low, approximately $1.86 \times 10^{-19}$.
We use Masstree~\cite{masstree} as the index in the \mn, which maps keys to the logID in data nodes.


\subsection{\bf Overall Performance}
\label{sub:overall}

\begin{figure}[]
  \begin{center}
    \includegraphics[width=\linewidth]{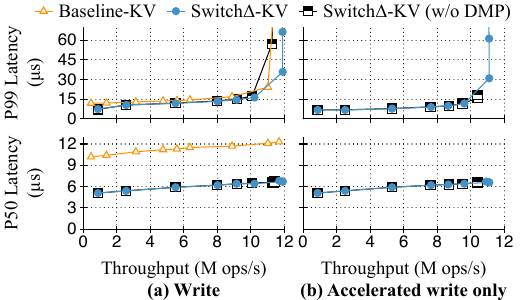}
  \end{center}
  \vspace{0.1cm}
  \mycaption{fig:kvbase}{P50 and P99 latency against throughput under a \textbf{write-only workload}}{(a) shows the performance of all operations, while (b) shows the performance of only the accelerated write operations in \sys-based systems.}
\end{figure}

\begin{figure}[]
  \begin{center}
    \includegraphics[width=\linewidth]{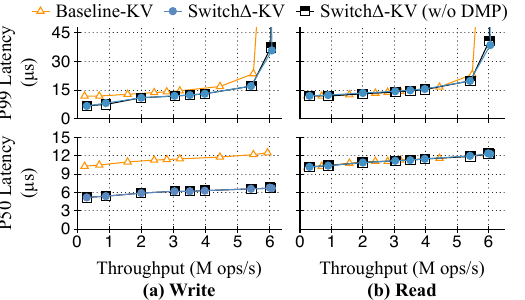}
  \end{center}
  \vspace{0.1cm}
  \mycaption{fig:kvbase2}{P50 and P99 latency for read and write operations under a \textbf{50\%/50\% read/write workload}}{(a) displays write performance, while (b) displays read performance.}
\end{figure}

We evaluate the performance of three systems in the same framework: Baseline-KV, \sys-KV, and \sys-KV (w/o DMP) which only employs in-switch data visibility protocol and disables Deferred Metadata Processing technology.
Fig.\ref{fig:kvbase} shows the P99 and median (P50) latency against the throughput as we increase the number of concurrent operations (i.e., the sum of clients' queue depths) from 6 to 768.
The workload in this experiment is write-only.

We make the following three observations:

First, \sys-KV reduces the median latency of write operations by 43.3\%$\sim$50.0\% compared to the Baseline-KV
This is because that \sys moves the data visibility phase out of the critical path, and the latency of write operations is only 1 RTT, the saved time includes the network between the switch and \mns and the queuing and processing time in \mns. 
In this experiment, the median latency of this saved period is between about 4.9$\sim$5.6$\mu s$, which is about half of the median write latency (10.1$\sim$12.3$\mu s$) in the baseline. 
Note that this improvement depends on the latency breakdown of the data write phase and the data visibility phase of a write operation. 
We evaluate performance under different configurations in \S\ref{sub:sensitive} and evaluate the performance when write operations have a heavy data write phase (i.e., data replication) in \S\ref{sub:rep}.


Second, \sys-KV reduces the 99th percentile latency of write operations by 39.4\% compared to the Baseline-KV, when the throughput is low (i.e., $<2Mops/s$, the first two points of each line in the figure). 
This is because, under this workload pressure, the count of concurrent operations is small, and almost all write operations (above 99.3\%) can be accelerated by the in-switch data visibility protocol. 


Third, the peak throughput of \sys-KV (w/o DMP) is similar to that of the Baseline-KV, and when limiting the 99th percentile latency bellowing 60$\mu s$, \sys-KV improves the peak throughput by 8.0\% compared to other two systems. 
The improvement comes from that: DMP saves CPU resources of \mns via the batching optimizations on these outside critical requests.
We give an in-depth evaluation of batching optimizations in \S\ref{subsub:batcheval}.



Fig.\ref{fig:kvbase2} shows the P99 and median latency of write operations and read operations, respectively, under the write-heavy workload (i.e., 50\% operations are write). 
In this experiment, we focus only on the performance of read operations. 
We observe that \sys does not impact read operations, both positively and negatively. 
On the one hand, only a very small fraction of metadata requests for read operations can be accelerated by the switch.
In this experiment, at most 4.7\% of the read operations are accelerated (total P50 read latency is 10.1$\mu s$$\sim$12.3$\mu s$, the P50 latency of the accelerated read operations is 7.3$\mu s$$\sim$10.4$\mu s$), and thus overall the P50 and P99 latency is not reduced. 
On the other hand, DMP only defers requests that are not on the critical path, whereas all requests for read operations are located on the critical path, and thus \sys does not add latency to read operations.


\subsection{\bf Sensitivity Analysis}
\label{sub:sensitive}

\begin{figure}[]
  \begin{center}
    \includegraphics[width=0.97\linewidth]{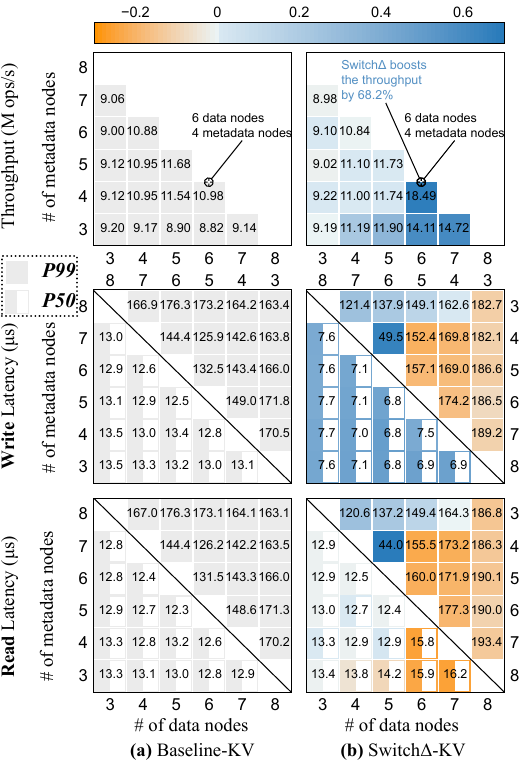}
    \mycaption{fig:threads}{Experiments on varying the numbers of \dn (X Axis) and the numbers of \mn (Y Axis)}{The figures indicate the throughput, P50 latency, and P99 latency. The colors indicate the performance positive (Blue, 0$\sim$70\%) and negative (Orange, -30\%$\sim$0) impacts from \sys. P99 latency corresponds to the \textit{right and top} Axis of the Figure.}
  \end{center}
\end{figure} 
To understand how \sys-KV performs under different resource allocations and identify its benefits when system bottlenecks shift, we conduct an experiment varying the number of data and metadata nodes. Fig.\ref{fig:threads} presents the results of this sensitivity analysis, showing how throughput and latency change as we scale the number of data nodes (from 3 to 8) and metadata nodes (from 3 to 8). All other settings, such as client and workload configurations, follow the default setup described in \S\ref{exsetup}.

\textbf{The key takeaway is that \sys~consistently reduces latency, and additionally boosts throughput when metadata processing is the bottleneck.} Specifically, when data nodes limit the system (the number of \dns is smaller than 6), \sys~focuses on reducing write latency by up to 49.2\%. However, when metadata processing becomes the bottleneck, \sys~not only reduces latency but also boosts overall throughput by up to 68.2\%.
We now detail these observations:

First, when Baseline-KV and \sys-KV achieve a similar throughput (i.e., when the number of \dns is less than 6), \sys reduces the P50 latency of write operations by 41.0\%$\sim$49.2\% and does not impact the P50 latency of read operations. This observation is consistent with \S\ref{sub:overall}.

Second, when the number of \dns is 6 or more (e.g., configurations (6,3), (6,4), and (7,3)), \sys boosts the throughput by 59.8\%$\sim$68.2\%. This is because the metadata processing (i.e., data visibility phase) becomes the throughput bottleneck. Our deferred metadata processing technology saves CPU resources on \mns, allowing them to process more metadata requests. The higher P50 read latency observed in these high-throughput configurations is a consequence of increased client-side pressure.

Third, when Baseline-KV and \sys-KV achieve a similar throughput, \sys-KV has a higher 99th percentile latency of all operations. This behavior is an artifact of our closed-loop evaluation model; because \sys-KV reduces the latency of most operations, clients can issue requests at a higher rate, which leads to increased contention and a higher probability of in-switch hash collisions. As a result, the percentage of accelerated write operations falls below 99\%. The 99th percentile latency then reflects the processing latency of the non-accelerated operations, which includes both the data write and data visibility phases, similar to the baseline. We evaluate how the number of concurrent operations and the workload skewness affect the percentage of accelerated write operations in \S\ref{sub:fastread}.

\subsection{\bf Workloads with long Data Write Phase}
\label{sub:rep}

The main contribution of in-network data visibility protocol is to move the latency of the data visibility phase outside the critical path. 
However, this contribution depends on the latency breakdown of the data write phase and the data visibility phase of a write operation. 
Fig.\ref{fig:rep} evaluates the performance of \sys with heavy data write phase (i.e., data replication).
In this experiment, we use a 3-way primary-backup replication configuration. When a data node receives a data write request, it uses one-sided RMDA WRITE to write its 2 backup replicas, and after receiving 1 ACK from the backup replicas, it installs its local log and replies to the client. Data nodes use coroutine to pipeline multiple data write requests. 
In our experiment, the data replication brings 3.6$\mu s$$\sim$4.0$\mu s$ latency overhead to the data write phase, leading to the benefits of in-network data visibility protocol reduce from 44.7\% to 30.0\%. As for throughput, since the data replication makes the data nodes the system bottleneck, \sys~does not show a performance improvement in this configuration, which is expected as \sys's core contribution is latency reduction.

\subsection{\bf In-depth Analysis of Each Technique}
\label{sub:indepth}

\subsubsection{In-Network Data Visibility Protocol.}

\label{sub:fastread}

\begin{figure}[]
  \begin{center}
    \includegraphics[width=\linewidth]{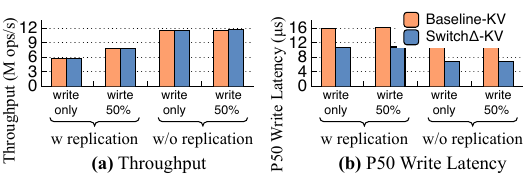}
    \vspace{-0.2cm}
    \mycaption{fig:rep}{Performance with 3-way replication on data}{}
  \end{center}
\end{figure} 

\begin{figure}[]
  \begin{center}
    \includegraphics[width=\linewidth]{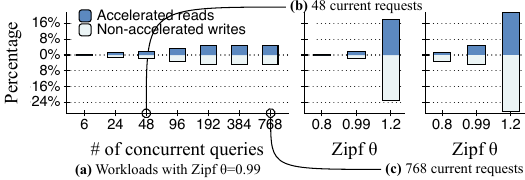}
    \mycaption{fig:overiw}{Percentage of non-accelerated write operations and accelerated read operations under different workloads}{(a) shows the effect of the number of concurrent requests, while (b) and (c) show the effect of workload skewness for 48 and 768 concurrent requests, respectively.}
  \end{center}
  
\end{figure}

In \sys, the switch handles read operations corresponding to its stored metadata. While this design is for ensuring strict consistency of read operations, it may also expedite these read operations. We term them as \textit{accelerated read operations}.
Due to hash collisions within the switch, some write operations cannot be stored in the switch and must instead follow the standard path, which involves two ordered RPCs. We term these write operations as \textit{non-accelerated write operations}.

We evaluate them under a write-heavy workload (50\% operations are write). We observe that the percentages of these two operations both fluctuate based on two factors: the number of concurrent requests and the skewness of the key distribution.

\noindent
\textit{a) Number of Concurrent Requests.}
Fig.\ref{fig:overiw}.a shows the percentage of accelerated read operations and non-accelerated write operations as we increase the number of concurrent operations from 6 to 768. 
The percentage of accelerated read operations is calculated as the number of read operations that hit the cache in the switch divided by the total number of read operations, and the percentage of non-accelerated write operations is calculated as the number of write operations that can not be stored in the switch divided by the total number of write operations.
The percentage of accelerated read increases from 0.2\% to 5.0\%, while the percentage of non-accelerated write operations increases from 0.2\% to 4.7\%.

\noindent
\textit{b) Skewness of Key Distribution.} 
We evaluate these percentages under the workloads with different Zipf $\theta$ parameters in Fig.\ref{fig:overiw}.b and Fig.\ref{fig:overiw}.c. 
These two experiments were performed, with a different number of concurrent requests, 48 and 768, respectively. 
As the Zipf $\theta$ increases from 0.8 to 1.2, 
the percentages of accelerated read operations and non-accelerated write operations increase up to 28.5\% and 21.4\%, respectively.
Note that, $\theta$=0.99 is a common test configuration.
With the configuration of 250M key space in our experiments, 49.1\% and 87.4\% of read operations access the hottest 1\textpertenthousand~ keys with the Zipf $\theta$ parameter of 0.99 and 1.2, respectively.

These results are consistent with theoretical expectations. We postulate a constant residence time for each metadata in the switch, i.e., the latency of the processing delay of the metadata node and the network communication delay between the switch and the metadata node.  
As the number of requests increases or as the key distribution becomes more skewed, the probability of queries and modifications on this metadata during its stay in the switch also rises.

Additionally, we observe that the percentages of these two operations are roughly equal in our experiments, this is because the ratio of read operations to write operations is 1:1 in the workloads. Once the metadata is installed in the switch, subsequent read operations on that metadata are accelerated, and subsequent write operations on that metadata fall back to the slower path.




\begin{figure}[]
  \begin{center}
    \includegraphics[width=\linewidth]{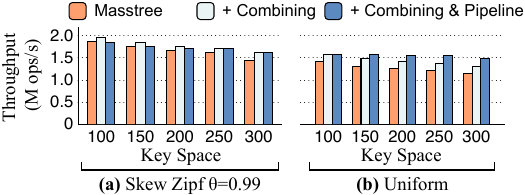}
  \end{center}
  \vspace{0.1cm}
  \mycaption{fig:key spacebatch}{Throughput of Batching Optimizations}{}
\end{figure}

\subsubsection{Deferred Metadata Processing.~}

\label{subsub:batcheval}
We evaluate the throughput of metadata requests under with different skewness (varying the Zipf distribution parameter $  \theta$), and with different key space configurations (100$\sim$300 Million) to show the benefits of batching optimizations. The batch size is 16.


Fig.\ref{fig:key spacebatch} shows the performance of metadata updates, since we use batching optimizations for write operations, which only require updates on the index.
We make the following observations: with the increase of the key space and with the decreases of Zipf $\theta$, the throughput improvement of batching optimizations increases from 4.7\% to 13.4\%. 
This is because the batching optimizations on the index improve throughput via reducing the CPU stall time for L3 cache misses. Under the workloads with larger key space, the index occupies more memory space and has poorer space localization; and under the workloads with uniform key distribution, the index operations have poorer time localization. Therefore, batching optimizations have more opportunities to improve the throughput under these workloads.

Further, the prefetching pipeline is a negative optimization when the key space is smaller than 200 million and the Zipf $\theta$ is larger than 0.99. 
This is because these workloads have few cache misses, i.e., less opportunity for using coroutine switching to hide the CPU stall time of cache misses. The coroutine switching requires the CPU to save and restore the context, which introduces additional overhead.

\subsection{\bf Recovery Time of Switch Failure.}
\label{subsec:eval:recovery}
We emulated a switch crash by performing a hard reboot. The system required 56 seconds to recover, about 32 seconds spent on re-initializing server-to-server connections and deploying the switch configuration, while 24 seconds spent on rebuild the metadata for 250M objects. During this recovery period, all ongoing client operations failed and were automatically retried until the network fabric was restored.

\section{\bf Case Studies}

\label{sec:app}

In each of the following sections, we show the basic architecture of the application first,
and describe the modification we made to the application to use \sys secondly.
The experiments show that \sys improves up to 126.9\% throughput and reduces up to 52.4\% latency of these applications.

\begin{figure*}[]
  \begin{center}
    \includegraphics[width=\linewidth]{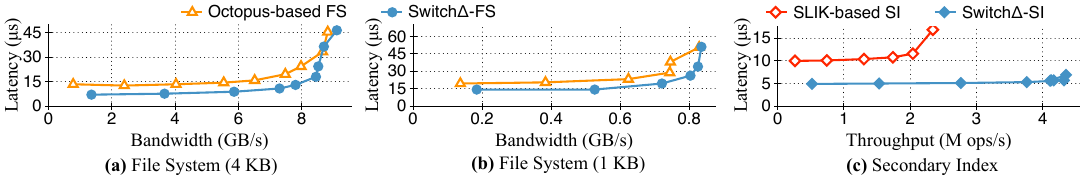}
  \end{center}
  \mycaption{fig:case}{Case Studies of Distributed File System and Distributed Secondary Index}{}
\end{figure*}

\subsection{\bf Distributed File System}

\subsubsection{Implementation of \sys-FS}

\label{sec:xx} 

We implement a distributed file system based on an open-source system, Octopus~\cite{octopus}. We store the metadata and the data on different nodes. In this experiment, we use 5 nodes (a data node, a metadata node, and 3 clients). The \dn stores the data blocks, which are 4KB and serve as the basic data unit of the file system. The \mn manages the directory and file inodes via a chained hash table, which uses the full path name as the key. Each inode stores the basic information of the file or directory (such as the file size and timestamps), the number of blocks in the file, and the addresses of these blocks.


A write operation in the distributed file system consists of three phases: (1) the client fetches the metadata of the file; (2) the client sends the write request with the metadata to the \dn, and the \dn uses a copy-on-write method to update the data blocks and returns the information of the new blocks; (3) the \mn updates the metadata to make the new modified blocks visible to all clients. For 4KB aligned writes, the file system allocates and writes the new block directly, skipping the first phase of reading the metadata. With \sys, the file system makes the write operation visible in the switch and moves the third phase out of the critical path.

\subsubsection{Evaluation.}
In this experiment, each client selects a file under Zipf distribution with a skewness of 0.99 in its own directory and reads/writes the file randomly. The directory number is equal to the client number, and each directory can have up to 32 files. We run the experiments with 5 nodes (a data node, a metadata node and 3 client nodes). We evaluate the median latency against the bandwidth as we increase the number of concurrent clients. Fig.\ref{fig:case}.a shows the results for 4KB aligned access, and Fig.\ref{fig:case}.b shows the results for 1KB access. We observe that \sys reduces the latency by up to 47.7\% and 28.2\% for 4KB and 1KB access, respectively. Under the 4KB-aligned access workload, \sys exhibits better performance improvement than that under the 1KB access workload. This is because the 4KB-aligned access operations skip the first phase of reading the metadata, making the benefits of removing the third phase more pronounced. We also observe that \sys does not improve the peak throughput of the file systems; this is because the bottleneck of the file system is the bandwidth of the \dn.

\subsubsection{Comparison with Other File Systems}
\label{rsec:fs}
The data-metadata separation architecture is a common design pattern adopted by many prominent distributed file systems, such as CephFS~\cite{ceph}, InfiniFS~\cite{infinifs}, 3FS~\cite{a3fs}, and HopsFS~\cite{hopsFS}, which share a similar architectural model to our work. For our evaluation, we chose Octopus~\cite{octopus} as the primary baseline. This decision was deliberate: Octopus is an open-source file system that leverages a high-performance RPC framework (RDMA-based). Crucially, its network stack is analogous to our user-space implementation, as both operate entirely in user-space and employ a polling-based model for communication. This choice facilitates a fair comparison, enabling us to precisely isolate and quantify both the relative and absolute performance gains achieved by offloading the ordered-write coordination to the switch. 
Furthermore, the principles of \sys are broadly applicable to the aforementioned file systems. They all can improve the write-performance by eliminating the metadata update phase from the critical path of write operations with an in-switch data visibility layer.

\subsection{\bf Distributed Secondary Index}
\label{sec:SI} 



\subsubsection{Implementation of \sys-SI}
In secondary index systems, the primary index and the secondary index are stored on different nodes, similar to SLIK~\cite{slik}. This approach allows independent partitioning and scaling of the primary index and the secondary index. The \dn stores the primary index, mapping the primary key to the address of the data. The \mn stores the secondary index, which combines the secondary key, timestamp, and primary key as a composite key mapped to the corresponding primary key. Both the primary index and the secondary index use Masstree, and we employ the range query of Masstree to implement the search operation of the secondary index.

The write operation updates or inserts a record to the primary indexes (i.e., the data write phase), inserts a new composite key into the secondary index (i.e., the data visibility phase) on the critical path, and deletes the old composite key in the secondary index in the background. With \sys, the secondary index system makes the write operation visible in the switch and moves the second phase out of the critical path. The read operation searches the secondary index to get the list of primary keys, searches the primary index to retrieve the records, and then validates them. It is important to note that the system without \sys also requires the validation step to support the background execution of the third phase. \sys repurposes the validation step to handle the hash collision problem, as described in \S\ref{subsubsec:hashcollision}.

\subsubsection{Evaluation.}
In this experiment, the key is 8-byte, and the timestamp is 4-byte. Therefore, the composite key in the secondary index is 20-byte. The key space of primary keys and secondary keys is 100 million and 4 million, respectively. The secondary key is randomly generated for each primary key, and on average, about 25 primary keys have the same secondary key. The operations are under the Zipf distribution with a skewness of 0.99. The read operations return the first 10 records on the target secondary key. We run the experiments with 5 nodes (a data node, a metadata node and 3 client nodes). 

Fig.\ref{fig:case}.c shows the median latency
against the throughput as we increase the number of concurrent operations. 
We observe that \sys improves the peak throughput by up to 81.1\% and reduces the latency under a few concurrent operations by up to 52.4\%. 
Compared with the performance improvement in the micro benchmarks, the performance improvement in the secondary index system is larger.
This is because the \mn needs to handle more operations in the secondary index system. 
Without \sys, the second phase is on the critical path and shares the resources with the third phase, thus the throughput is limited and the latency is also affected by queuing issues in the \mn. 

\subsubsection{Comparison with Other Secondary Index Systems}
\label{rsec:si}
Rocksteady~\cite{kulkarni2017rocksteady} provides seminal work on dynamic index partitioning and rescheduling to handle load imbalances. Its architecture, which separates primary and secondary indexes, also validates the prevalence of the data-metadata separation model that we target. Our work is orthogonal: while Rocksteady addresses the control plane problem of data placement and load balancing, \sys focuses on accelerating the data plane's "ordered write" protocol. Integrating \sys with such dynamic systems by adapting our in-switch routing to software-level changes is a promising direction for future work, making our solution more robust in highly elastic environments.


\section{\bf Related Work}

\subsection{\bf Metadata Cache in Distributed Systems}

\label{rsec:cache}
It is instructive to compare \sys with alternative caching strategies, such as client-side caching and other in-switch caching mechanisms.

Client-side caching, as seen in systems like Ceph or Google File System, presents a fundamental trade-off. To maintain data consistency across multiple clients, these systems must either employ complex and costly cache coherence protocols or relax consistency guarantees. The latter often requires modifying the application's design, for instance, by enforcing append-only data models to avoid in-place updates, or by providing weaker semantics like close-to-open, where data visibility is only guaranteed after an explicit flush. In contrast, \sys offers a solution that balances both strong consistency and high performance without requiring any changes to the upper-level application semantics.

In-switch caching has been used to store hot data for high-performance operations (NetCache~\cite{netcache} and P4DB~\cite{p4db}), offload control-plane metadata (NetLock~\cite{netlock}, Concordia~\cite{concordia}, SwitchTx~\cite{switchtx}, and CableCache~\cite{hotstorage25cablecache}), or store transient operational state like global sequencers~\cite{Eris,1pipe,hydra}. \sys~is fundamentally different. Instead of being a conventional cache, it uses the switch as a temporary buffer to decouple the two phases of the ordered write, thus accelerating the write path. Entries are allocated for in-flight operations and immediately reclaimed upon completion, which obviates the need for complex cache management policies (e.g., LRU) and makes our design highly efficient for resource-constrained switches.

\subsection{\bf Asynchronous Metadata Updating}
ScaleDB~\cite{scaledb} employs a fast, small hash index to manage delayed updates and asynchronously updates the range index. Additionally, some log-structured storage systems~\cite{F2FS, NOVA, boyd2014oplog, Logstructured, lsm, leelog} promptly acknowledge write operations immediately after appending data to the log and updating the in-memory index. They primarily target the sequential-write-friendly storage devices to keep random writes out of the critical path.

The core principle of \sys, utilizing a simple yet efficient layer (i.e., the programmable switch) to absorb metadata updates and applying these updates to metadata nodes in the background, aligns with these systems. However, \sys is specifically designed for distributed systems, leveraging programmable switches to redesign the network workflow of operations. Furthermore, unlike ScaleDB, \sys derives performance benefits not only from the performance disparities between the hash index and range index but also from reduced network communication.

\subsection{In-Switch Acceleration Techniques}
Accelerating systems with P4 switches requires navigating hardware limitations in memory and expressiveness. Our work builds on prior in-switch techniques, particularly in data structures and consistency. A key challenge is implementing dynamic data structures; while match-action tables are suitable for static keys~\cite{netcache}, register arrays are preferred for dynamic state~\cite{ask, farreach, switchhash, hotstorage25cablecache}. Like others, \sys~adopts a hash-based index to support large keys. However, while some systems avoid collisions by pre-assigning hashes for static data~\cite{netrpc}, \sys~handles dynamic workloads via a lightweight, runtime validation and retry mechanism. To ensure consistency, \sys~adopts a routing-based model, where all operations must traverse the switch to observe the latest state, similar to NetCache~\cite{netcache} and FarReach~\cite{farreach}. This contrasts with invalidation-based protocols, such as the one used by DistCache.

\section{\bf Conclusion}

This paper introduces \sys, an in-s witch data visibility layer that accelerates distributed storage systems. By buffering in-flight metadata updates in the switch, \sys~makes metadata updates asynchronous, which can halve write latency. Furthermore, it leverages this asynchrony to enable batch processing, boosting throughput without impacting user-perceived latency. Our evaluations show that \sys~yields significant performance improvements across various storage systems, especially under write-heavy workloads.

\section*{Acknowledgements}
We sincerely thank anonymous reviewers for their valuable feedback, which significantly improved this paper. This work is supported by the National Natural Science Foundation of China (Grant No. 62332011, 62472242), Beijing Natural Science Foundation (Grant No. L242016), Huawei and Young Elite Scientists Sponsorship Program by CAST (2023QNRC001).

\bibliographystyle{unsrt}
\bibliography{sample.bib}

\end{document}